\documentclass[aps,prb,twocolumn,showpacs]{revtex4-1}
\pdfoutput=1
\usepackage{graphicx}
\usepackage{dcolumn}
\usepackage{bm}
\usepackage{amsmath}
\usepackage{color}
\usepackage[normalem]{ulem}

\begin{document}

\newcommand{\bra}[1]{\left\langle#1\right|}
\newcommand{\ket}[1]{\left|#1\right\rangle}
\newcommand{\bracket}[2]{\big\langle#1 \bigm| #2\big\rangle}
\newcommand{\PPV}{\ket{0_{\rm PP}}}
\newcommand{\Tr}{{\rm Tr}}
\renewcommand{\Im}{{\rm Im}}
\renewcommand{\Re}{{\rm Re}}
\newcommand{\MC}[1]{\mathcal{#1}}
\newcommand{\pp}{{\prime\prime}}
\newcommand{\ppp}{{\prime\prime\prime}}
\newcommand{\pppp}{{\prime\prime\prime\prime}}

\title{Orbital signatures of Fano-Kondo line shapes in STM adatom spectroscopy}

\author{Sebastian Frank} 
\affiliation{Max-Planck-Institut f\"ur Mikrostrukturphysik, Weinberg 2, 06120 Halle, Germany}
\author{David Jacob} 
\affiliation{Max-Planck-Institut f\"ur Mikrostrukturphysik, Weinberg 2, 06120 Halle, Germany}

\date{\today} 

\begin{abstract}
  We investigate the orbital origin of the Fano-Kondo line shapes measured
  in STM spectroscopy of magnetic adatoms on metal substrates. To this end we 
  calculate the low-bias tunnel spectra of a Co adatom on the (001) and (111) 
  Cu surfaces with our density functional theory-based \emph{ab initio} transport scheme augmented by 
  local correlations. In order to associate different $d$-orbitals with different 
  Fano line shapes we only correlate individual $3d$-orbitals instead of the full 
  Co $3d$-shell. We find that Kondo peaks arising in different $d$-levels indeed 
  give rise to different Fano features in the conductance spectra. 
  Hence the shape of measured Fano features allows us to draw some conclusions
  about the orbital responsible for the Kondo resonance, although the actual shape
  is also influenced by temperature, effective interaction and charge fluctuations.  
  Comparison with a simplified model shows that line shapes are mostly the result of 
  interference between tunneling paths through the correlated $d$-orbital and the 
  $sp$-type orbitals on the Co atom. 
  Very importantly, the amplitudes of the Fano features vary strongly among orbitals,
  with the $3z^2$-orbital featuring by far the largest amplitude due to its strong
  direct coupling to the $s$-type conduction electrons.
\end{abstract}

\maketitle

\section{Introduction}
\label{sec:intro}

The Kondo effect is one of the most fascinating phenomena in condensed matter physics,
occurring in a vast number of different systems 
(see, e.g., Ref.~\onlinecite{Hewson1997} and references therein), ranging from bulk metals 
doped with magnetic impurities\cite{deHaas1934,Sarachik1964,Kondo1964} to nanoscale systems 
such as semiconductor quantum dots\cite{Posazhennikova2007,Roch2008} and carbon nanotubes 
connected to metal leads\cite{Nygard2000,Jarillo-Herrero2005}. 
Generally, the Kondo effect leads to the quenching of a local magnetic moment associated with 
localized and strongly interacting electronic states in the system by interaction with the 
conduction electrons. The quenching of the spin is accompanied by drastic changes in 
the electronic and transport properties. This strong impact on the electronic
and magnetic properties of a system makes the Kondo effect an important factor 
for the functionality of atomic and molecular-scale electronic devices.

Since the pioneering works of Li {\it et al.}\cite{Li1998} and Madhavan {\it et al.}\cite{Madhavan1998} 
scanning tunneling spectroscopy (STS) has become a standard tool for probing the Kondo effect of magnetic 
adatoms and molecules on metallic substrates\cite{Manoharan2000,Knorr2002,Nagaoka2002,Heinrich2004,Zhao2005,Crommie2005,Iancu2006,Neel2007}.
The Kondo effect arises from the interaction of the magnetic moment of
the adsorbate with the conduction electrons of the metal surface, and leads to the screening of the magnetic 
moment by formation of a total spin singlet state with the conduction electrons.
The formation of the Kondo-singlet state is signaled by the appearance of a strongly 
renormalized quasi-particle peak at the Fermi level, the so-called Abrikosov-Suhl or Kondo resonance. 
In STS the appearance of the Kondo peak in the local DOS of the atom or molecule leads to a 
zero-bias anomaly (ZBA) in the tunnel spectra which is generally well described by a Fano line shape\cite{Fano1961}, 
although it has recently been found that the ZBAs are actually much better described in terms of generalized
Frota line shapes\cite{Prueser2012} as the Frota function yields a much better description of the Kondo peak
than the Lorentz function.\cite{Frota1986,Frota1992}

The origin of the Fano-like line shape is either understood as 
due to the interference of different tunneling paths - one via the 
strongly interacting orbitals of the magnetic atom bearing the 
sharp Kondo resonance, and others going directly to the substrate\cite{Plihal2001,Madhavan2001,Lin2006} 
- or is explained in terms of tunneling into the surface alone\cite{Ujsaghy2000,Schiller2000,Wahl2004,Merino2004a,Merino2004b}. 
A recent study\cite{Baruselli2015} combines density functional theory (DFT) with numerical 
renormalization group (NRG) calculations and determines the line shape by looking 
at energy-dependent transmission eigenvalues.
Surprisingly, no systematic study of the relation between orbital
symmetry of the orbital(s) bearing the Kondo resonance and the shape of 
the resulting Fano resonances has been conducted so far.

In this paper, we intend to close this gap by calculating the 
Fano line shapes corresponding to Kondo peaks appearing in different 
orbitals of the $3d$-shell of a magnetic atom on metal surfaces.
To this end we select individual $d$-orbitals and perform \emph{ab initio} 
quantum transport calculations augmented by local correlations for the 
selected $d$-orbital only. There is merit in doing so: Even in a multi-orbital 
situation the Kondo effect is signaled by Kondo peaks in individual $d$-orbitals, 
and often the Fano-Kondo feature of one $d$-orbital will be dominant in the tunnel 
spectrum due to different tunneling matrix elements and Kondo scales.

We choose to study Co$@$Cu(001) 
and Co$@$Cu(111) as our test systems, which have been extensively studied theoretically
and experimentally\cite{Vitali2008,Neel2010,Merino2004a,Merino2004b,Choi2012,Knorr2002,Neel2007,Wahl2004,Lin2006,Baruselli2015}. 
We find that Kondo peaks arising in different
$d$-levels indeed give rise to different Fano features in the conductance
spectra. However, temperature, effective interaction and occupancy of the $d$-orbital 
also play an important role. With one notable exception, a simplified two-level model 
consisting of the $d$-orbital bearing the Kondo resonance and one $s$- or $p$-orbital
on the adatom accounts for the calculated line shapes. This shows that in these cases
tunneling into substrate states only plays a minor role for determining the actual line shapes.

The paper is organized as follows: In Sec.~\ref{sec_method} we briefly describe
the method for calculating the zero-bias anomalies in the conductance spectra 
corresponding to Kondo peaks in different
orbitals. In Sec.~\ref{sec_fano} we introduce two types of Fano line shapes: the standard one
based on a Lorentzian resonance for the localized state and one based on the Frota
line shape better suited for describing the Kondo resonance.
In Sec.~\ref{sec_results} we present results for a Co adatom placed on a Cu(001) and a Cu(111) 
surface, respectively. In Sec.~\ref{sec_models} we devise a simplified model capturing the essence of the 
different situations encountered for different orbital symmetries and discuss the obtained 
results in the context of this model. In Sec.~\ref{sec_discussion}, a more general discussion follows
relating our results to other experimental and theoretical works.
Finally, in Sec.~\ref{sec_conclusions}, we conclude this work with 
some general remarks on the significance of our results for other atomic 
or molecular Kondo systems.

\section{Method}
\label{sec_method}

\begin{figure}
  \begin{center}
    \includegraphics[width=.22\textwidth]{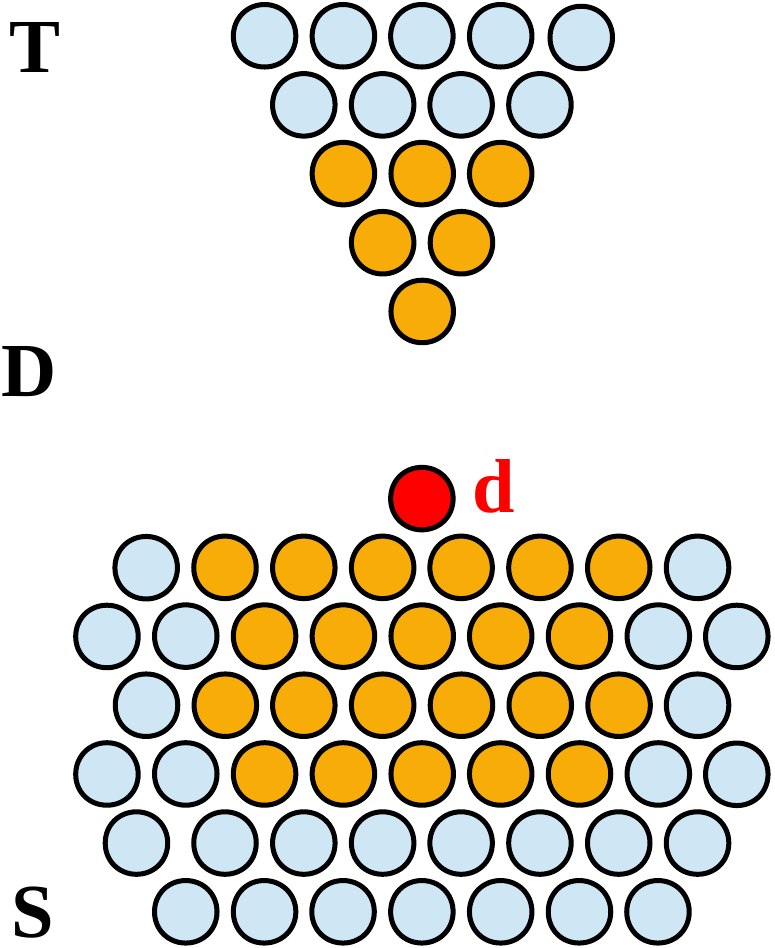} 
  \end{center}
  \caption{
    Schematic drawing of an STM tip probing a magnetic
    atom on a metal substrate. The system is separated into
    three parts: the device region D (gray/yellow) contains 
    the magnetic atom (dark gray/red) hosting the $d$-orbital giving
    rise to the Kondo peak, and parts of the substrate and STM 
    tip. T and S (light gray/blue) are the bulk electrodes connected
    to the STM tip and substrate, respectively.
  }
  \label{fig_method_system}
\end{figure}

We consider a magnetic atom (here: Co) that is placed on a metallic substrate (here: Cu(001) or Cu(111)). 
A Cu STM tip is placed directly above the Co atom 6\r{A} away so that we are in the tunneling
regime. The system is divided into three parts as shown in Fig.~\ref{fig_method_system}: two metal leads S and 
T, representing the bulk electrodes connected to the substrate and STM tip, respectively, and the device region 
(D) which contains the magnetic atom and part of the surface and the STM tip.

We perform DFT based \emph{ab initio} quantum transport calculations
using the ANT.G package\cite{Jacob2011}: The electronic structure of the D region is
calculated on the level of Kohn-Sham (KS) DFT employing the LSDA functional\cite{Kohn1965} 
in the SVWN parametrization \cite{Slater1974,Vosko1980} and a minimal Gaussian basis set 
including the valence (4s4p3d) and outer core electrons (3s3p) of the Co and Cu 
atoms~\cite{Hehre1969,Collins1976,Hay1985a,Wadt1985,Hay1985b}.
The effect of the bulk electrodes S and T, which are modeled by Bethe lattices\cite{Bethe1935},
on the electronic structure of D is taken into  account via self-energies $\hat{\Sigma}_{\rm S}$ and 
$\hat{\Sigma}_{\rm T}$. The KS Green's function (GF) of the D region is thus given by:
\begin{equation}
  \hat{G}^0_{\rm D}(\omega) = \left( (\omega+\mu)\hat{P}_{\rm D} - \hat{H}_{\rm D}^0 
  -\hat{\Sigma}_{\rm T}(\omega) -\hat{\Sigma}_{\rm S}(\omega)\right)^{-1} \textrm{,} 
\end{equation}
where $\mu$ is the chemical potential, $\hat{P}_{\rm D}$ the projection operator onto D
and $\hat{H}_{\rm D}^0$ is the KS Hamiltonian of the D region.

In order to capture Kondo physics, electronic correlation beyond conventional DFT have to be included. 
This is done by combining DFT with the one-crossing approximation (OCA)\cite{Haule2001}, following 
the scheme developed in previous work~\cite{Jacob2009,Jacob2015}. In contrast to previous work, 
we are interested in the Kondo signatures of specific $d$-orbitals, and not of the entire $3d$-shell. 

Hence we add a Hubbard-like interaction term $\hat{\mathcal{H}}_{U} = U \hat{n}_{d \uparrow} \hat{n}_{d \downarrow}$ only 
to a \emph{single} $d$-orbital of the Co $3d$-shell where $n_{d\sigma}$ is the number operator for the 
$d$-orbital and a spin $\sigma$. 
Since the Coulomb interaction in the correlated $d$-orbital has already been taken into account on 
a mean-field level in the KS-DFT calculation, a double-counting correction (DCC) term has to be subtracted
from the KS Hamiltonian projected onto the $d$-orbital $\epsilon_d^0=\bra{d}\hat{H}_{\rm D}^0\ket{d}$: 
\begin{equation}
  \epsilon_{d} = \epsilon^0_{d} - \epsilon_{\rm dc} \textrm{.}
\end{equation}
In contrast to previous work the DCC is chosen such that a certain occupancy is achieved,
i.e. for achieving particle-hole (ph) symmetry ($n_d=1$) we choose $\epsilon_{\rm dc}$ such that
$\epsilon_{d}=-U/2$. Note that ph symmetry is only approximately achieved since the coupling
of the $d$-orbital to the rest of the system (see below) is generally not ph symmetric.

The interacting $d$-orbital coupled to the electronic bath given by the rest of the system 
(i.e. substrate and tip) defines an Anderson impurity model (AIM)\cite{Anderson1961}. An 
effective description of the coupling of the $d$-orbital to the bath is given by the so-called 
hybridization function $\Delta_d(\omega)$ which can be obtained from the KS GF by
\begin{equation}
  \Delta_d(\omega) = \omega + \mu - \epsilon^0_{d} - \left[ G^0_d(\omega) \right]^{-1} \textrm{,}
\end{equation}
where $G^0_d$ is the KS GF projected onto the $d$-orbital, i.e. $G^0_d=\bra{d}\hat{G}^0_{\rm D}\ket{d}$.
The imaginary part of $\Delta_d$ yields the broadening $\Gamma_d$ of the $d$-orbital due to
the coupling to the rest of the system.

The AIM is now solved in the OCA.\cite{Haule2001} 
The solution yields the self-energy $\Sigma_{d}(\omega)$ describing the dynamic 
correlations of the $d$-orbital. The correlated GF of the $d$-orbital is then given by 
\begin{equation}
G_d(\omega) = \left( [G_d^0(\omega)]^{-1} - \Sigma_d(\omega) + \epsilon_{\rm dc} \right)^{-1} \textrm{.}
\end{equation}
Its imaginary part yields the spectral function or LDOS of the $d$-orbital $\rho_d(\omega)=-\Im{}G_d(\omega)/\pi$.
Correspondingly, we obtain the correlated GF for the D region as:
\begin{equation}
  \hat{G}_{\rm D}(\omega) = \left( [\hat{G}_{\rm D}^0(\omega)]^{-1} -(\Sigma_d-\epsilon_{\rm dc})\hat{P}_d \right)^{-1} \textrm{.}
\end{equation}
This allows us to calculate the transmission function using the Caroli expression \cite{Caroli1971}, 
\begin{equation}
  T(\omega) = \Tr\left[ \hat{G}_{\rm D} \hat\Gamma_{\rm T} \hat{G}_{\rm D}^{\dagger} \hat\Gamma_{\rm S} \right]
  \label{eqn_caroli}
\end{equation}
where the coupling matrices for the leads are defined by
\begin{equation}
  \hat{\Gamma}_{\rm T/S} = i \left( \hat{\Sigma}_{\rm T/S} - \hat{\Sigma}_{\rm T/S}^{\dagger} \right) \textrm{.}
\end{equation}
The self energies $\Sigma_{\rm T/S}$ are typically symmetric, so that the coupling matrices are twice the imaginary 
part of the self energies. 
For low temperature and small bias voltages, current and conductance can be related to the transmission 
function using the Landauer formula\cite{Landauer1957,Datta1995}. For the typical STM setup considered here 
most of the applied bias voltage drops at the STM tip. In that case the conductance is simply given by
\begin{equation}
  \label{conductance}
  G(V) = \frac{2e^2}{h} T(eV) \textrm{.}
\end{equation}
We note that the use of the Landauer formula for the conductance is justified in the limit of 
small bias voltages compared to the Kondo temperature. In this limit transport occurs via the Kondo 
resonance and thus is essentially one-body like (apart from renormalization) and phase coherent so that
the full non-equilibrium expression for transport through an interacting region given by Meir-Wingreen 
reduces to the simpler Landauer result.\cite{Meir1992} For larger bias voltages, deviations from the 
Landauer result can occur\cite{Hettler1998,Balseiro2010,Roura-Bas2010}, 
and one would have to make use of the Meir-Wingreen equation\cite{Meir1992} which requires the solution of 
the AIM out of equilibrium.

\section{Fano-Lorentz and Fano-Frota line shapes}
\label{sec_fano}
Fano line shapes or resonances, originally introduced by Fano in the context of autoionization and
elastic electron scattering by helium\cite{Fano1961}, generally arise in resonant scattering processes due 
to quantum interference between a quasi discrete resonant state and a broad background continuum. 
The interference leads to an asymmetric line shape in the scattering cross section 
at energies close to the resonance energy
that is well described by the Fano function 
\begin{equation}
  f(\epsilon) \propto \frac{(q+\epsilon)^2}{\epsilon^2+1}  \textrm{,}
  \label{eqn_fano_original}
\end{equation}
where the parameter $q$ controls the shape of the Fano function, 
and $\epsilon$ is the energy with respect to the resonant level. 
Eq.~\ref{eqn_fano_original} can also be obtained from 
the complex representation of a Lorentzian multiplied 
by a phase factor $e^{i\phi_q}$
\begin{equation}
  \label{eqn_fano_lorentz}
  \rho_{\rm FL}(\omega) = \Im \left[ e^{i \phi_q} \left( \frac{A}{\omega - \omega_0 + i \Gamma} \right) \right] + \rho_0 \textrm{,}
\end{equation}
where $A$ is the amplitude, $\Gamma$ is the half-width of the Lorentzian,
$\omega_0$ the resonance energy and $\rho_0$ a constant offset. 
Using $q=\tan(\phi_q/2)$, $\epsilon = (\omega-\omega_0)/\Gamma$ and 
some algebra, this Fano-Lorentz (FL) line shape can be shown to be equivalent 
to the original Fano formula (see the Appendix):
\begin{equation}
  \label{eqn_fano_alternative}
  \rho_{\rm FL}(\omega) = \frac{A}{\Gamma} \left( \frac{(q+\epsilon)^2}{\epsilon^2+1} - 1 \right) \frac{1}{1+q^2} + \rho_0 \textrm{.}
\end{equation}

\begin{figure}
  \begin{center}
    \includegraphics[width=.99\linewidth]{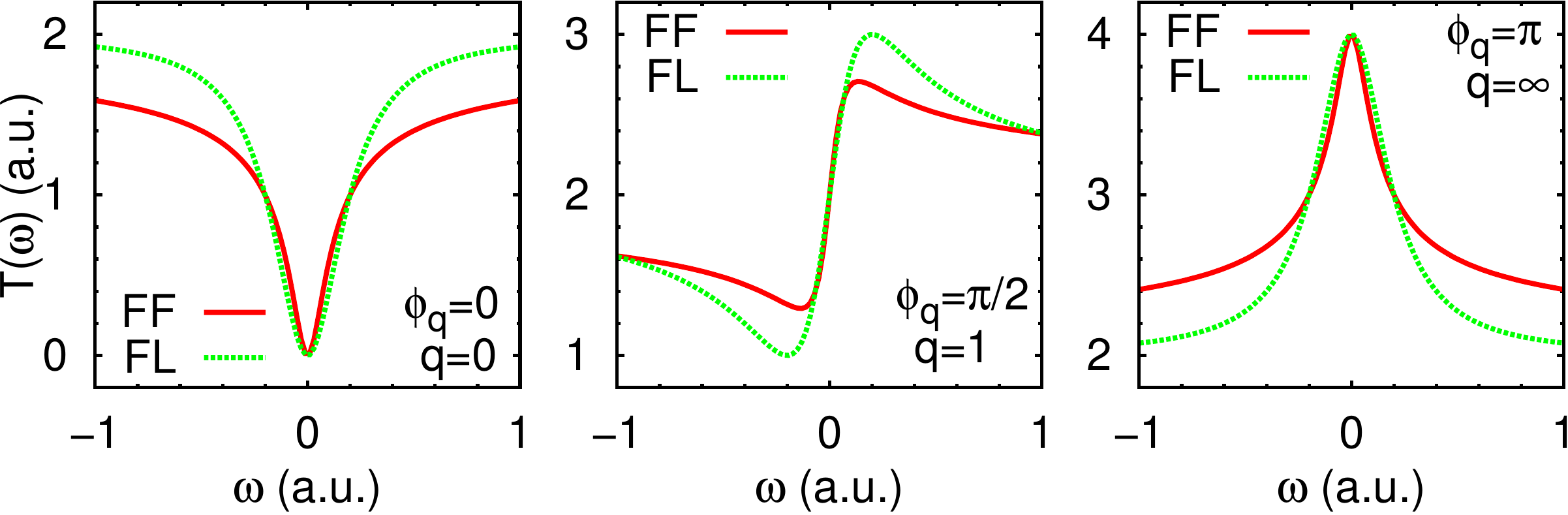} 
    \caption{Comparison of Fano-Frota and Fano-Lorentz line shapes for different values of the $q$ parameter
      but for identical amplitudes and half-widths.}
    \label{fig_frota_line shapes}
  \end{center}
\end{figure}

STM spectroscopy of Kondo impurities presents a similar situation: The STM tip probes 
the continuous conduction electron density of states which interacts with the Kondo resonance
at the Fermi level. The interference of different tunneling paths then leads to 
Fano-type line shapes in the conductance spectra. Assuming a Lorentzian form for the 
Kondo resonance naturally leads to Fano-Lorentz line shapes given by (\ref{eqn_fano_alternative}).
However, in Refs.~\onlinecite{Frota1986,Frota1992} Frota showed that the Kondo peak is 
actually better described by a line shape now known as a Frota line shape: 
\begin{equation}
  \label{eqn_frota}
  \rho_{\rm Frota}(\omega) =  A \cdot \Re\left[ \sqrt{\frac{i \Gamma_F}{\omega - \omega_0 + i \Gamma_F}} \right]
  \textrm{.}
\end{equation}
where the Frota parameter $\Gamma_{\rm F}$ is related to the actual half-width $\Gamma$ of the resonance
by $\Gamma\sim2.54\,\Gamma_{\rm F}$. $A$ is the amplitude and $\omega_0$ the position of the Frota resonance.
In analogy with Eq. (\ref{eqn_fano_lorentz}) we define a Fano-Frota (FF) line shape as a generalized
Frota curve~\cite{Prueser2011} for describing the transmission function close to the Kondo 
resonance
\begin{equation}
  \label{eqn_fano_frota}
  T_{\rm FF}(\omega) =  -A \cdot \Re\left[ e^{i \phi_q} \sqrt{\frac{i \Gamma_{\rm F}}{\omega - \omega_0 + i \Gamma_{\rm F}}} \right] + T_0  
  \textrm{.} 
\end{equation}
The phase $\phi_q$ has the same meaning in the Lorentz and in the Frota case: 
A value of $\phi_q=0$ leads to a dip, $\phi_q=\pi$ to a peak and $\phi_q=\pi/2$ to 
a symmetric Fano line shape. In Fig.~\ref{fig_frota_line shapes}, we compare 
Fano-Frota and Fano-Lorentz features, choosing identical amplitudes and half-widths. 
Note that for the same half-width Frota line shapes have a slower 
decay than the Lorentzian ones.

\begin{figure}
  \begin{center}                         
   \includegraphics[width=\linewidth]{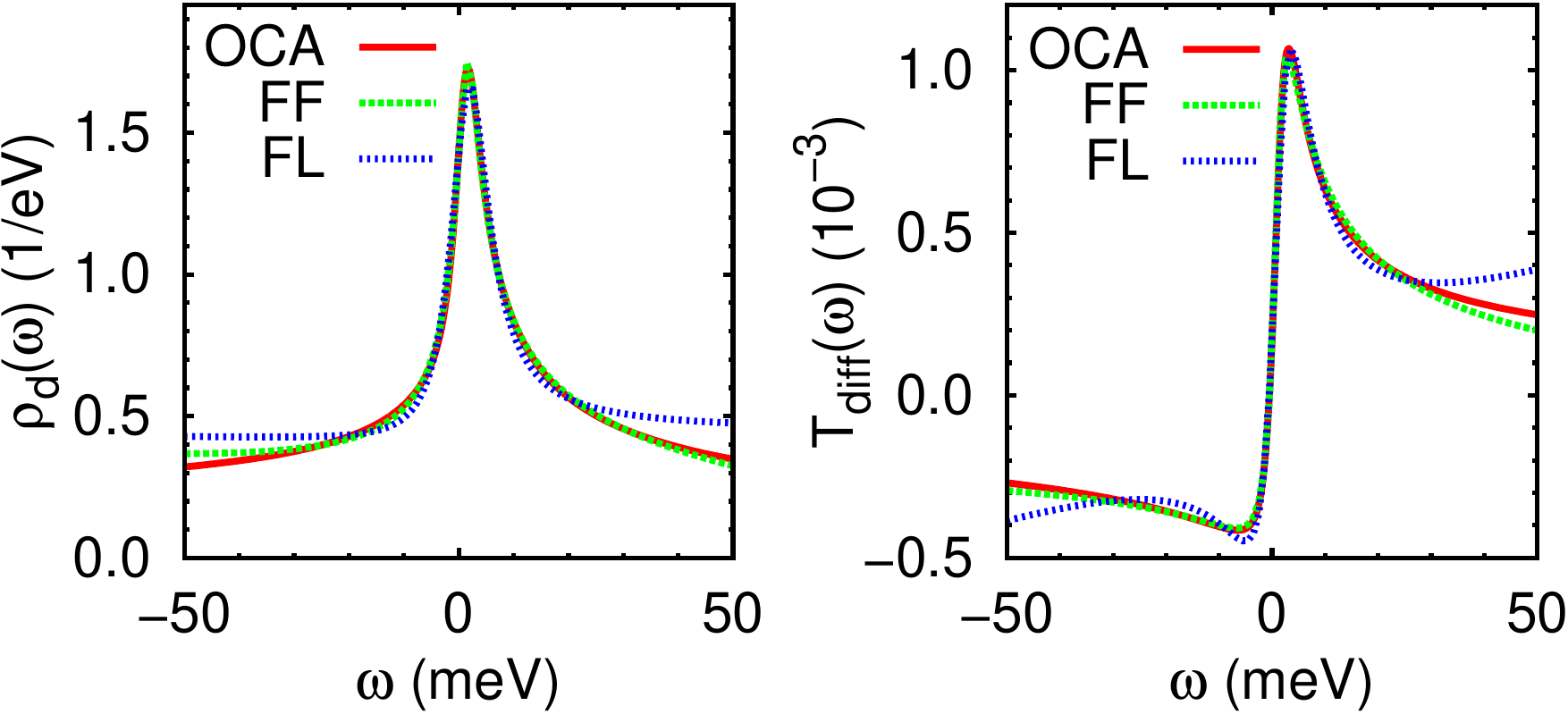}
   \caption{ 
     Fano-Lorentz (FL) and Fano-Frota (FF) fits of the impurity spectral function 
     $\rho_d(\omega)$ (left) and the transmission function $T(\omega)$ (right) 
     for $z^2$-orbital, Co@Cu(001), $U=2$~eV and $\epsilon_d=-1.0$~eV. See Sec.~\ref{sec_results_001}
     for details.
   }
   \label{fig_frota_fits}
  \end{center}
 \end{figure}

In Fig.~\ref{fig_frota_fits}, we show FL and FF fits to the Kondo peak (left) in the calculated spectral function $\rho_d(\omega)$ 
and the corresponding Fano line shape (right) in the calculated transmission function $T(\omega)$ for the case of the $z^2$-orbital
for the Co on Cu(001) system, discussed in detail in the following section. For both spectral and 
transmission function, the resonance center is well-described by FF and FL fits. However, only the FF fit yields an 
accurate description of the flanks and the long range decay. In the following, we will therefore use Eqn.~\ref{eqn_fano_frota}
to fit transmission functions. 

\section{Results}
\label{sec_results}

\subsection{Co adatom on Cu(001) surface}
\label{sec_results_001}

\begin{figure}
\begin{center}
 \includegraphics[width=.205\textwidth]{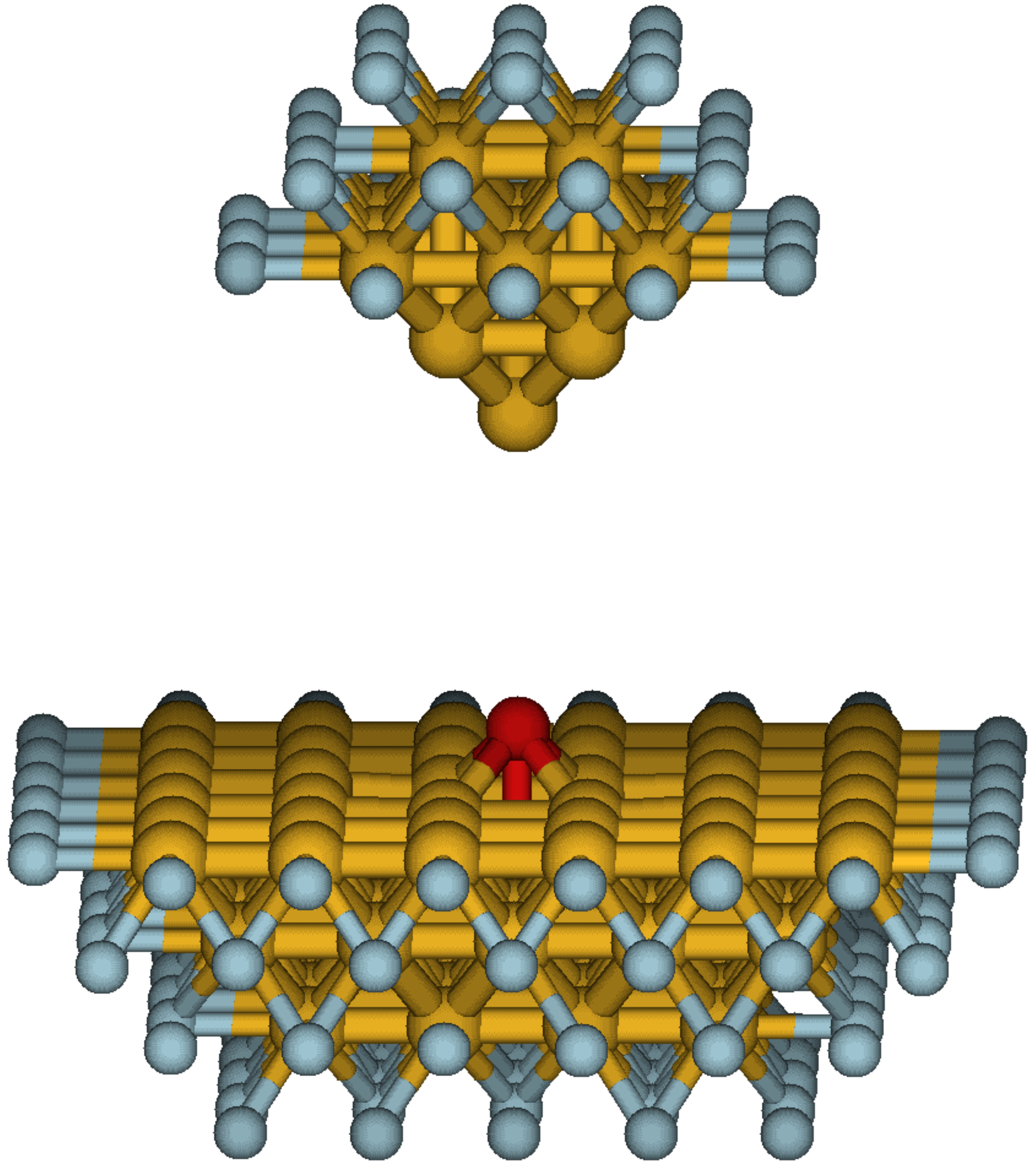} \quad 
 \includegraphics[width=.253\textwidth]{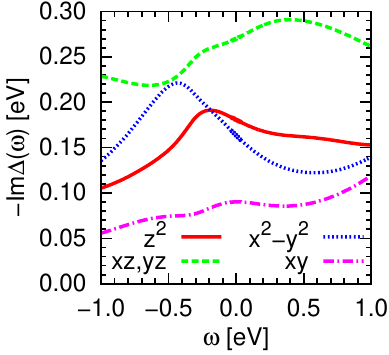}
 \caption{Left: Geometry of the Co atom deposited on a Cu(001) surface; 
   dark gray/red: Co, gray/yellow: Cu, light gray/blue: Bethe lattice. 
   Right: Imaginary part of the hybridization function for the Co $3d$ shell.}
 \label{fig_001_geohyb}
 \end{center}
\end{figure}

The system under consideration is shown in the left panel of Fig~\ref{fig_001_geohyb}. 
A Cobalt atom is deposited at the hollow site of 
a Cu(001) surface. The Cu(001) surface is modeled by three Cu slabs of 36, 25  and 16 atoms, respectively, 
which are embedded into a Bethe lattice to describe the infinitely  extended surface. 
We model the STM tip by a small pyramid of Cu atoms grown in the (001) direction, also embedded into a Bethe lattice. 
The tip is placed directly above the Co atom in a distance of 6\AA, so that the system is in the tunneling 
regime. 

As explained in Sec.~\ref{sec_method} we now compute the hybridization functions
of the Co $3d$-orbitals (see right panel of Fig.~\ref{fig_001_geohyb}). The four-fold
symmetry of the Cu(001) surface leads to a splitting into four groups.
The $xz$- and $yz$-orbitals are degenerate (in the following, results for the 
$yz$-orbital are omitted) and exhibit the strongest hybridization 
at the Fermi level. 
The hybridization functions of $z^2$ and $x^2$-$y^2$ have comparable values around the
Fermi level. The $xy$-orbital has the lowest hybridization in the displayed energy window. 
All hybridization functions show a moderate energy dependence.
Note that the hopping between different Co $3d$ orbitals is zero, i.e. they do not couple
to each other on the single-particle level.

While the hybridization function is calculated \emph{ab initio}, the Coulomb interaction 
$U$ is used as a parameter that allows us to tune the Kondo coupling strength and explore the 
effect of the width of the Kondo peak on the transmission line shape. But in order to have
an estimate of the magnitude, we have also calculated $U$ \emph{ab initio} for each of the 
$d$-orbitals by constrained RPA calculations as described in Ref.~\cite{Jacob2015}. 
We find values for $U$ ranging from 1.8~eV to 2.6~eV.\footnote{
  More specifically, we obtain 1.80~eV for the $z^2$-orbital, 1.78~eV for the $xz$- and $yz$-orbitals, 
  1.81~eV for the $x^2-y^2$ orbital and 2.59~eV for the $xy$ orbital. 
}
Accordingly, we choose the $U$ parameters to vary between 2~eV and 3~eV.

The hybridization functions from Fig.~\ref{fig_001_geohyb} together with the
energy level $\epsilon_d$ and the effective Coulomb interaction $U$ define 
an AIM which is solved in the OCA~\cite{Haule2001}. It is a known issue of OCA that at too
low temperatures (1-2 orders of magnitude below $T_K$) it gives rise to spurious
non-Fermi liquid behavior and related artifacts in the impurity spectral 
function, leading to an overestimation of the height of the Kondo peak and
an unphysical self-energy with positive imaginary part~\cite{Costi1996}. 
We circumvent this problem by lowering the temperature only to the point 
where the imaginary part of the self-energy becomes zero. At this point 
Fermi liquid behavior is obeyed, and the unitary limit of the Kondo peak 
is exactly recovered. 

\begin{figure}
  \begin{center}
    \includegraphics[width=.45\textwidth]{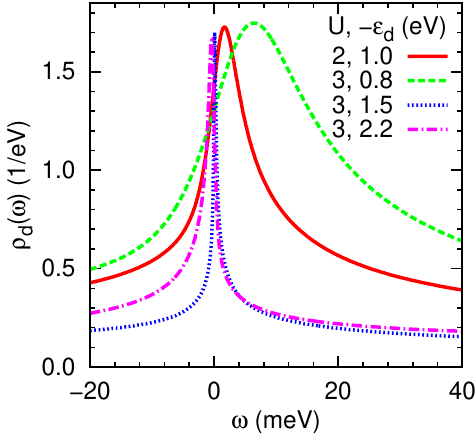} 
    \caption{Impurity spectral functions for the $z^2$ orbital of Co$@$Cu(001) 
             for different Anderson impurity model parameters $U$, $\epsilon_d$. }
    \label{fig_001_Aloc}
  \end{center}
\end{figure} 
 
Fig. \ref{fig_001_Aloc} shows impurity spectral functions $\rho_d(\omega)$ of the $z^2$-orbital
for different values of the AIM parameters $\epsilon_d$ and $U$.  
For $\epsilon_d=-U/2$ (red solid and blue dotted curves) we have approximate particle-hole
symmetry: the Kondo peak is centered close to, but slightly above the Fermi level. 
Note that exact particle-hole symmetry is not achieved because of the non-constant 
hybridization function. As expected, when $U$ is increased the Kondo temperature and hence
the width of the Kondo peak decrease strongly. On the other hand detuning the system
from particle-hole symmetry by shifting $\epsilon_d$ leads to a strong increase of the 
Kondo temperature due to charge fluctuations (green dashed, magenta dashed-dotted curves).

\begin{figure}
  \begin{center}
    \includegraphics[width=\linewidth]{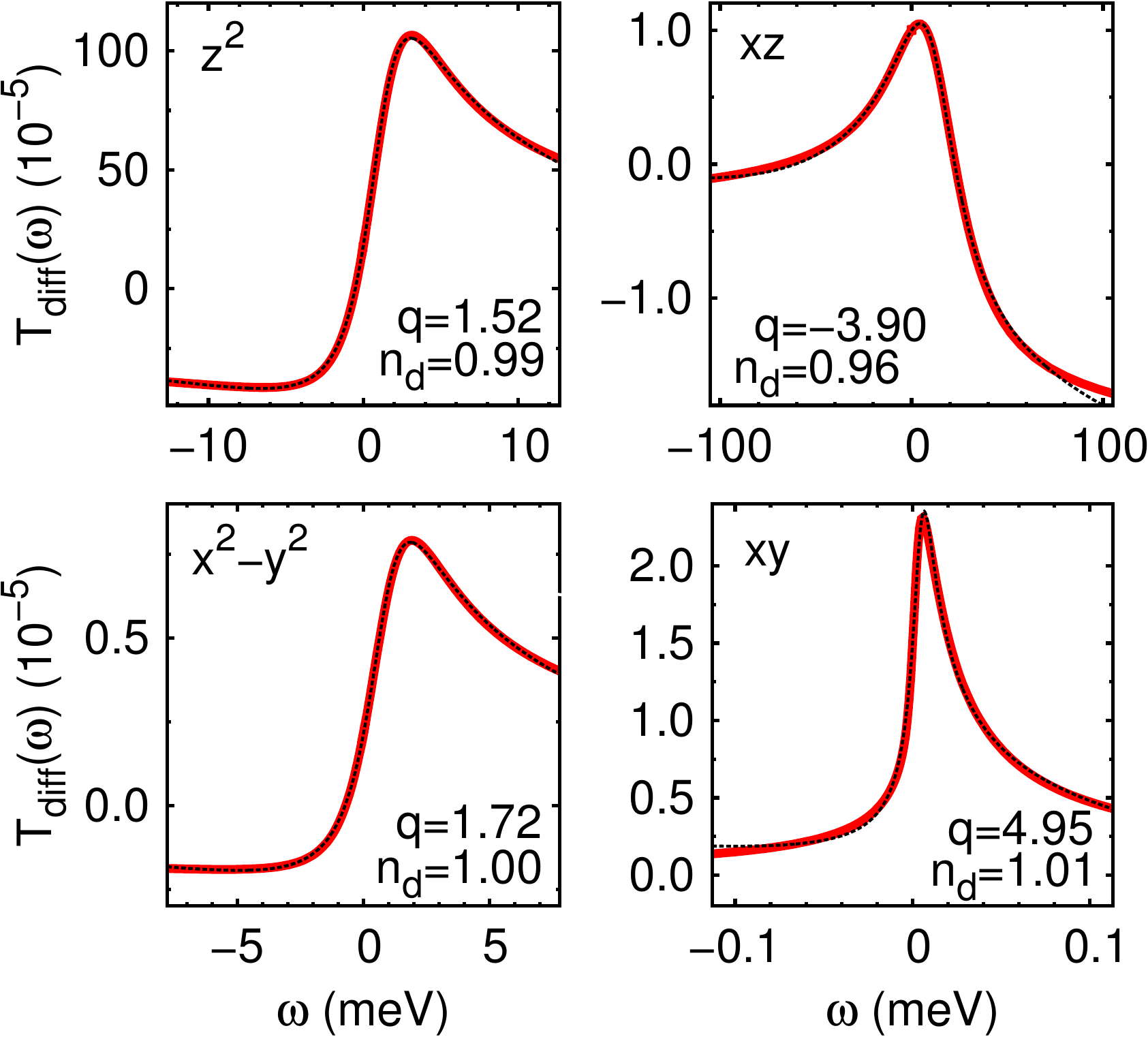}
    \caption{Transmission functions for different $d$-orbitals of Co$@${}Cu(001). 
      Coulomb repulsion $U=2$~eV, and energy level $\epsilon_d=-1$~eV (approximate particle-hole symmetry). 
      The red continuous curves show the calculated transmission, the black dashed curves Fano-Frota fits. 
      The transmission background has been subtracted.\cite{Note2} 
    }
    \label{fig_001_U2}
  \end{center}
\end{figure}

We now calculate the correlated transmission functions for Kondo peaks in different $d$-orbitals. 
Fig.~\ref{fig_001_U2} shows 
transmission line shapes for different $d$-orbitals for $U=2$~eV and \mbox{$\epsilon_d=-1.0$~eV}. 
In order to make the features more clearly visible, 
here and in the following the transmission background was subtracted.
\footnote{
  The background is calculated by calculating the transmission function
  without adding the self-energy but pushing the respective $d$~level 
  away from the Fermi level.
}
We find that the line shapes are indeed different for each orbital. 
We observe approximately antisymmetric Fano line shapes ($q\approx1$) 
for $z^2$ and $x^2-y^2$, and more peak-like feature ($q\gg1$) for $xz$ 
and $xy$. In order to quantitatively describe the line shapes, we 
perform Frota fits to determine the $q$ parameter and width of the line shapes,
as explained before in Sec.~\ref{sec_fano}. 
The $z^2$ and $x^2-y^2$ orbitals have comparable $q$ values of 1.52 and 1.72, respectively. 
For $xz$, $q$ becomes negative ($-3.9$) and for $xy$ we find the most pronounced peak with $q=4.95$. 
The width of the Fano features differs significantly, and in accordance with their hybridization 
strength at the Fermi level.
Note that a feature with a very small width, as e.g., in the case of $xy$, might never be 
observed in an actual experiment, because of the Kondo temperature being much too low 
and because of limited resolution.

\begin{figure}
  \begin{center}
    \includegraphics[width=\linewidth]{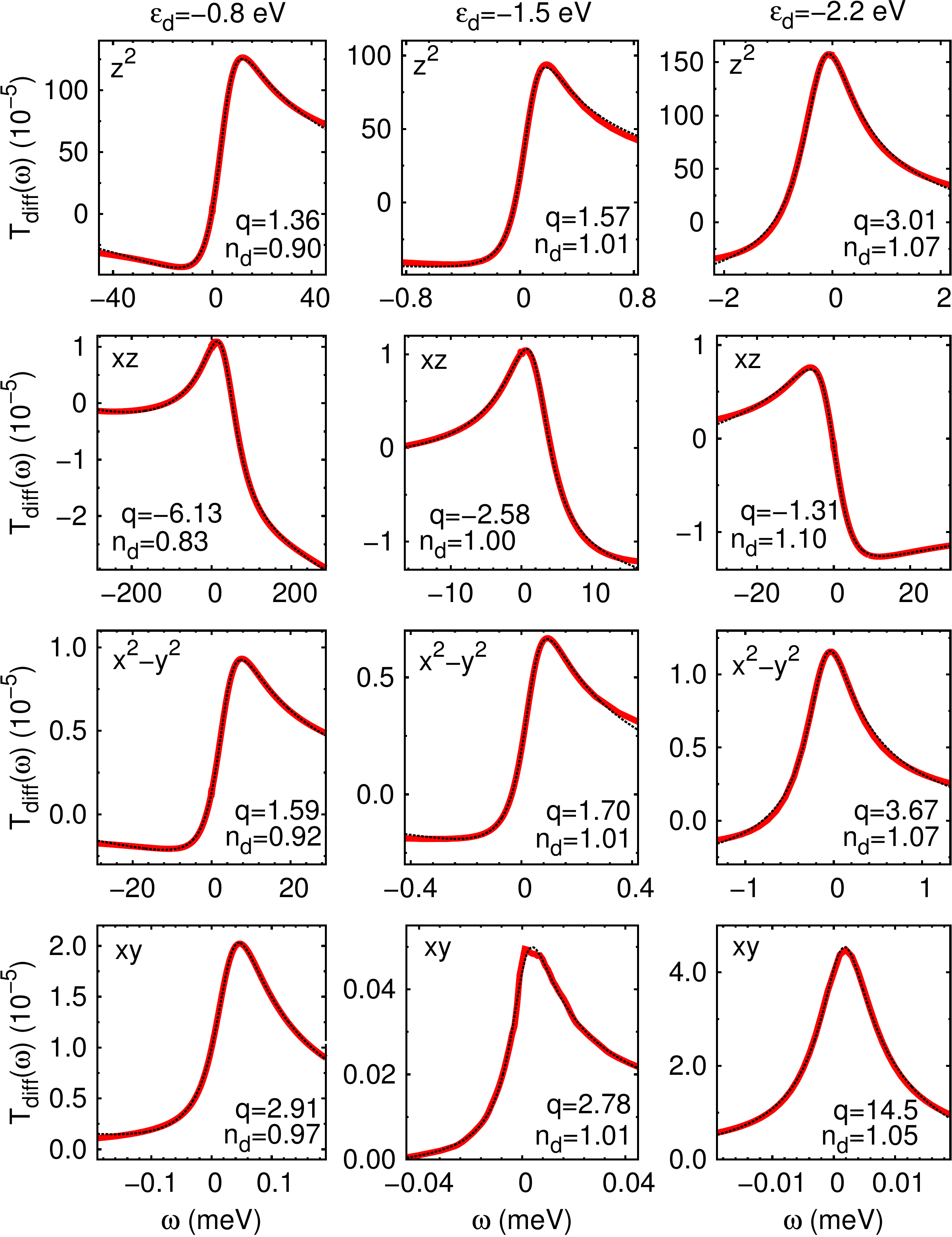} 
    \caption{ Transmission functions for different $d$-orbitals of Co$@${}Cu(001). Coulomb repulsion $U=3$~eV, 
      vary occupation by shifting $\epsilon_d$.  The red continuous curves show the calculated transmission, the black dashed curves Fano-Frota fits. 
      The transmission background has been subtracted.\cite{Note2} 
      \label{fig_001_U3}}
  \end{center}
\end{figure}

We now vary the Coulomb repulsion $U$ and introduce charge fluctuations by shifting the $d$-level 
position $\epsilon_d$, as can be seen in Fig.~\ref{fig_001_U3}. When varying $U$, but 
maintaining particle-hole symmetry, the actual shape of 
the transmission features is only weakly affected, while the widths of the features
change strongly, as has already been seen and discussed for the spectral functions in Fig.~\ref{fig_001_Aloc}. 
When introducing charge fluctuations, the Kondo peak becomes asymmetric (see Fig.~\ref{fig_001_Aloc}). 
This asymmetry is also reflected in the transmission line shapes. We find that the $q$ parameter consistently increases when $\epsilon_d$ is 
shifted downwards. For positive $q$ ($z^2$, $x^2-y^2$, $xy$) lowering $\epsilon_d$ makes the line shapes more peak-like, 
while for negative $q$ ($xz$), lowering $\epsilon_d$ leads to more dip-like line shapes. 

Hence while the choice of AIM parameters $U$ and $\epsilon_d$ does affect the transmission line shapes to some 
degree, it does not completely change its symmetry. For example, the sign of the $q$ factor does not change.

While the signal width is determined by the hybridization and choice of AIM parameters exclusively, the signal 
amplitude decisively depends on the system geometry. Because we chose the $z$-axis as our transport direction, a
Kondo peak in the $z^2$-orbital results in a much more dominant feature compared to the remaining $d$-orbitals, 
as can be seen in Figs.~\ref{fig_001_U2} and~\ref{fig_001_U3}. 
Hence if there is a Kondo peak in the $z^2$-orbital, the corresponding Fano feature will dominate in the 
transmission regardless of what happens in the other orbitals. Also Fano features due to Kondo peaks in orbitals 
other than the $z^2$-orbital might be difficult to discern from the background if the background dispersion is strong compared 
to the Fano amplitudes.
This statement remains true even if the STM tip is shifted laterally by moderate distances of a few \r{A}. 
Although tunneling into orbitals other than $z^2$ becomes more favorable upon a lateral shift of the tip, the feature due 
to the Kondo peak in the $z^2$ remains the most dominant one.

\subsection{Co adatom on Cu(111) surface}
\label{sec_results_111}

\begin{figure}
\begin{center}
 \includegraphics[width=.205\textwidth]{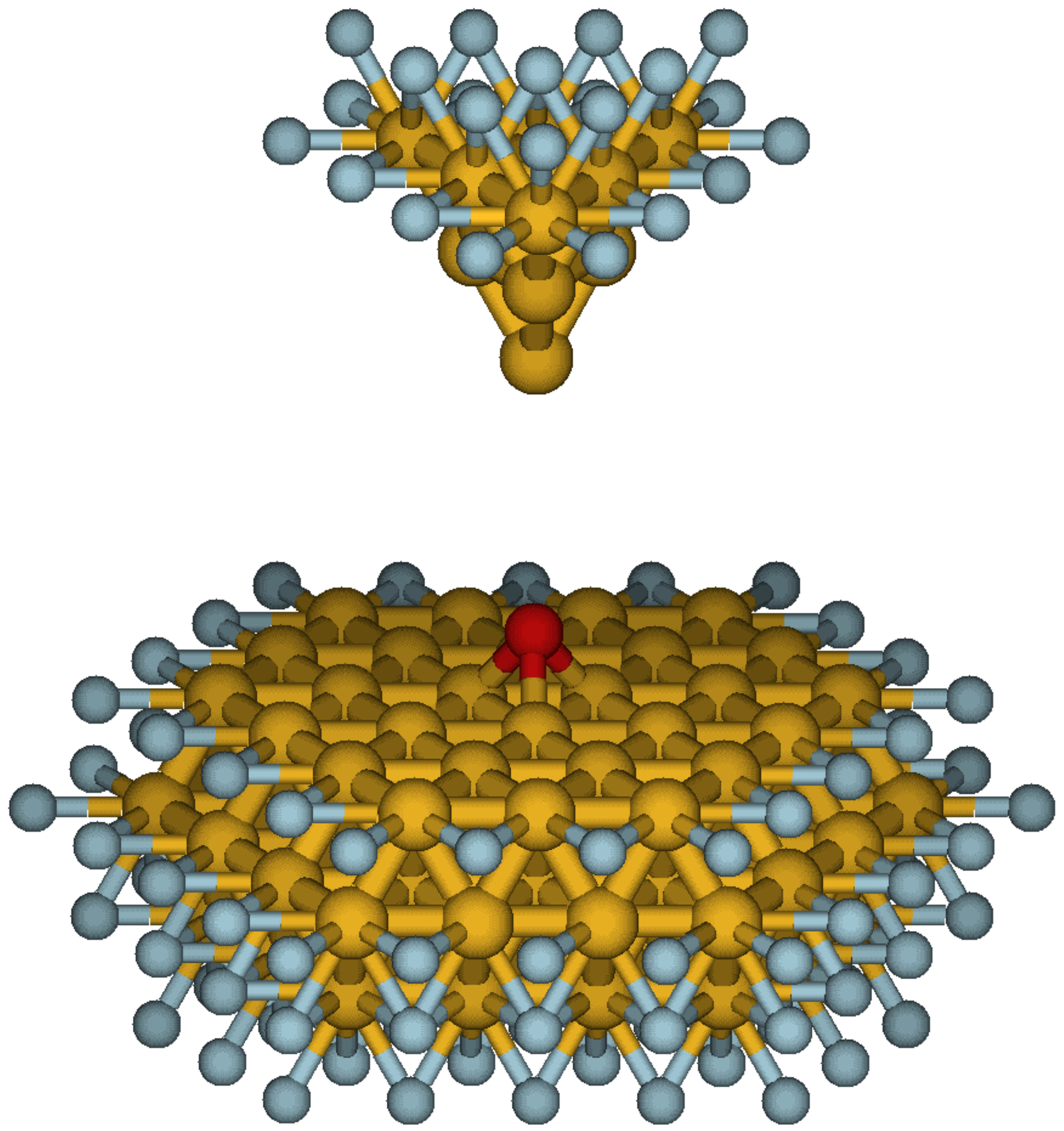} \quad 
 \includegraphics[width=.253\textwidth]{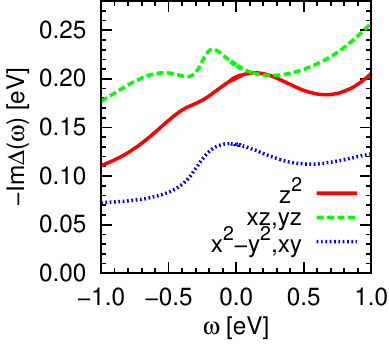}
 \caption{Left: Geometry of the Co atom deposited on a Cu(111) surface;
   dark gray/red: Co, gray/yellow: Cu, light gray/blue: Bethe lattice. . 
   Right: Imaginary part of the hybridization function of the Co $3d$-shell.}
 \label{fig_111_geohyb}
 \end{center}
\end{figure}

The next system we focus on is a Cobalt atom, deposited at the 'hcp' hollow site  
of a Cu(111) surface, as can be seen in the left panel of Fig.~\ref{fig_111_geohyb}. The surface is modeled by three 
Cu slabs of 27, 37 and 27 atoms, respectively, which are connected to a Bethe lattice. The tip is described by a Cu(111) 
pyramid, consisting of 10 copper atoms, also connected to a Bethe lattice. The threefold 
symmetry splits the five orbitals of the Co $3d$-shell into three groups: the non-degenerate $z^2$-orbital ($m=0$) and
two doubly degenerate groups, one with $m=\pm1$ ($xz$- and $yz$-orbitals) and one with $m=\pm2$ ($xy$- and $x^2-y^2$-orbitals).
The right panel of Fig.~\ref{fig_111_geohyb} shows the hybridization functions for each of the three groups.
The group with the $xz$- and $yz$-orbitals exhibit the strongest hybridization at the Fermi level, 
the group with the $x^2-y^2$- and $xy$-orbitals the weakest. 

\begin{figure}
 \begin{center}
  \includegraphics[width=\linewidth]{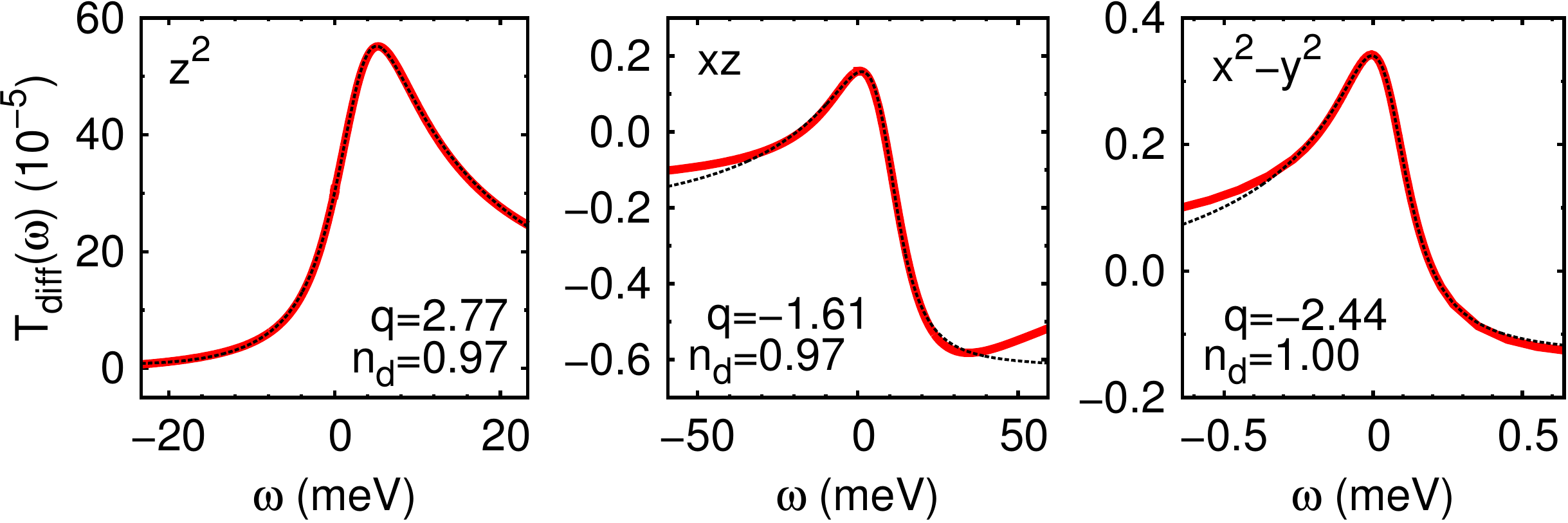} 
 \caption{Transmission functions for different $d$-orbitals of Co$@${}Cu(111). 
   Coulomb repulsion $U=2$~eV, $\epsilon_d=-1.0$~eV.
    The red continuous curves show the calculated transmission, the black dashed curves Fano-Frota fits. 
   The transmission background has been subtracted.\cite{Note2} 
 }
 \label{fig_111_U2}
 \end{center}
 \end{figure}

We proceed as described in the previous section and calculate transmission functions for the $d$-orbitals of 
Co$@${}Cu(111), assuming a Coulomb repulsion of $U=2~eV$ and (approximate) particle-hole-symmetry $\epsilon_d=-1.0$~eV 
(Fig.~\ref{fig_111_U2}).
Again, we find different line shapes for each orbital. The $z^2$ orbital gives the most peak-like transmission feature 
with $q=2.77$, for $x^2-y^2$ we observe a transmission peak with a negative $q=-2.44$. The 
$xz$-orbital results in a Fano-type feature with $q=-1.61$. The widths of the transmission features differ
considerably, with the $xz$- and $yz$-orbitals having the largest width, and the $xy$- and $x^2-y^2$-orbitals
the lowest. The $z^2$-orbital again has the highest signal amplitude, as it couples strongly to the tip conduction
electrons.

\begin{figure}
  \begin{center}
    \includegraphics[width=\linewidth]{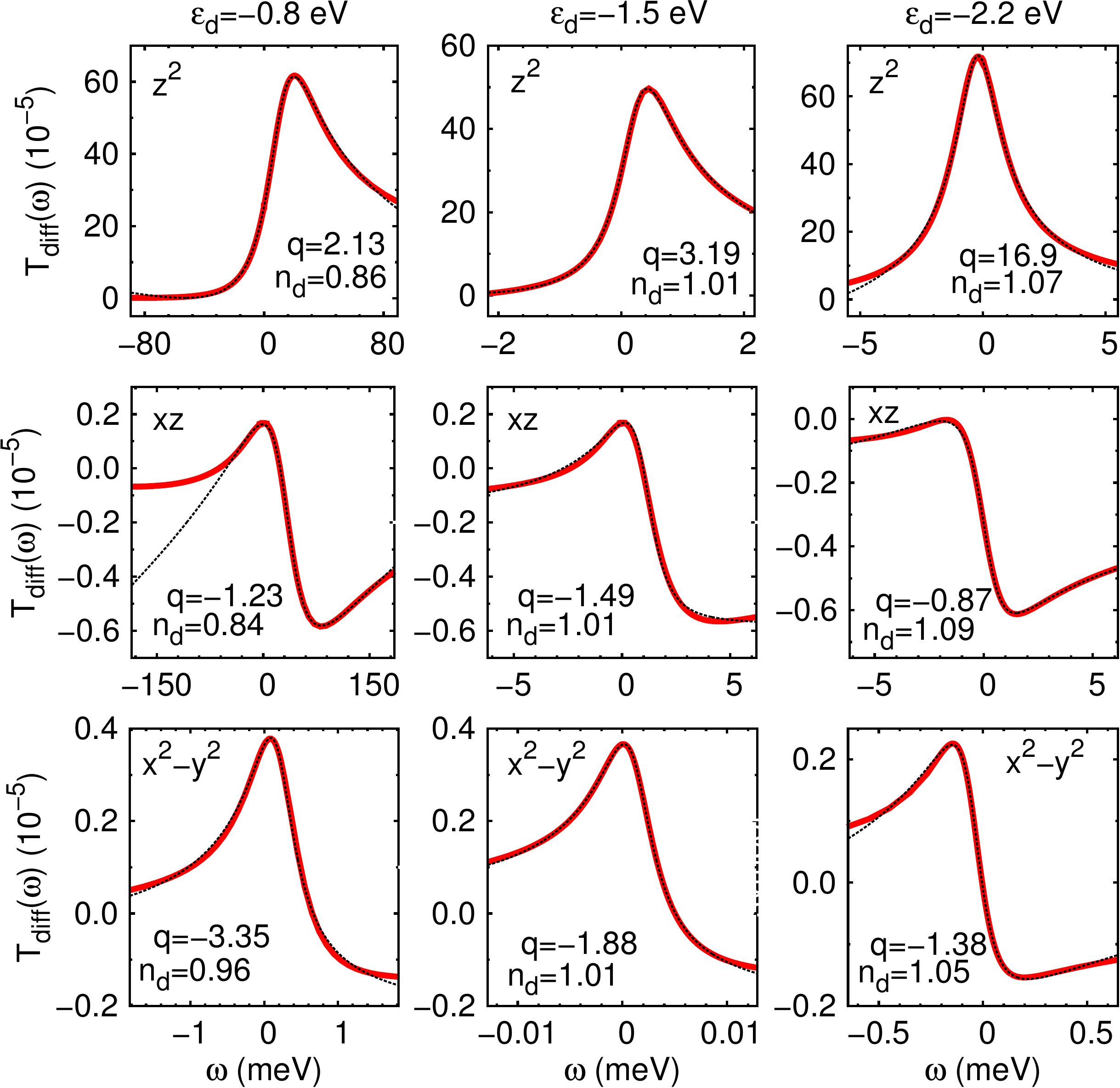}
    \caption{Transmission functions for different $d$-orbitals of Co$@${}Cu(111). 
      Coulomb repulsion $U=3$~eV, vary occupation by shifting $\epsilon_d$.
       The red continuous curves show the calculated transmission, the black dashed curves Fano-Frota fits. 
      The transmission background has been subtracted.\cite{Note2} 
    }
    \label{fig_111_U3}
  \end{center}
\end{figure}

In Fig.~\ref{fig_111_U3}, we calculate line shapes for different AIM parameters $U$ and $\epsilon_d$. 
We observe a similar behavior as for Co$@${}Cu(001). When staying in the particle-hole symmetric case and 
increasing $U$ (middle column of Fig.~\ref{fig_111_U3}), the line shapes remain similar, with slightly increased 
$q$ values. We introduce charge 
fluctuations by shifting the position of $\epsilon_d$ (left and right column of Fig.~\ref{fig_111_U3}). The $q$
parameter increases when moving $\epsilon_d$ to lower energies. For positive $q$ values, as for $z^2$, this leads
to more peak-like line shapes, while for negative $q$ values, as for $xz$ and $x^2-y^2$, it leads to more fano- or
dip-like line shapes. The only exception to this behavior occurs for the $xz$ orbital, $U=3$~eV 
and $\epsilon_d=-0.8~eV$. It has a very high Kondo temperature and equivalently wide Fano feature, and the Fano-Frota 
fit fails for negative energies. This suggests that the Fano line shape overlaps with other transmission features that 
alter the final line shape.
\subsection{Temperature dependence and the occurrence of dips}
\label{sec_results_Tdep}
\begin{figure}
  \begin{center}                         
   \includegraphics[width=\linewidth]{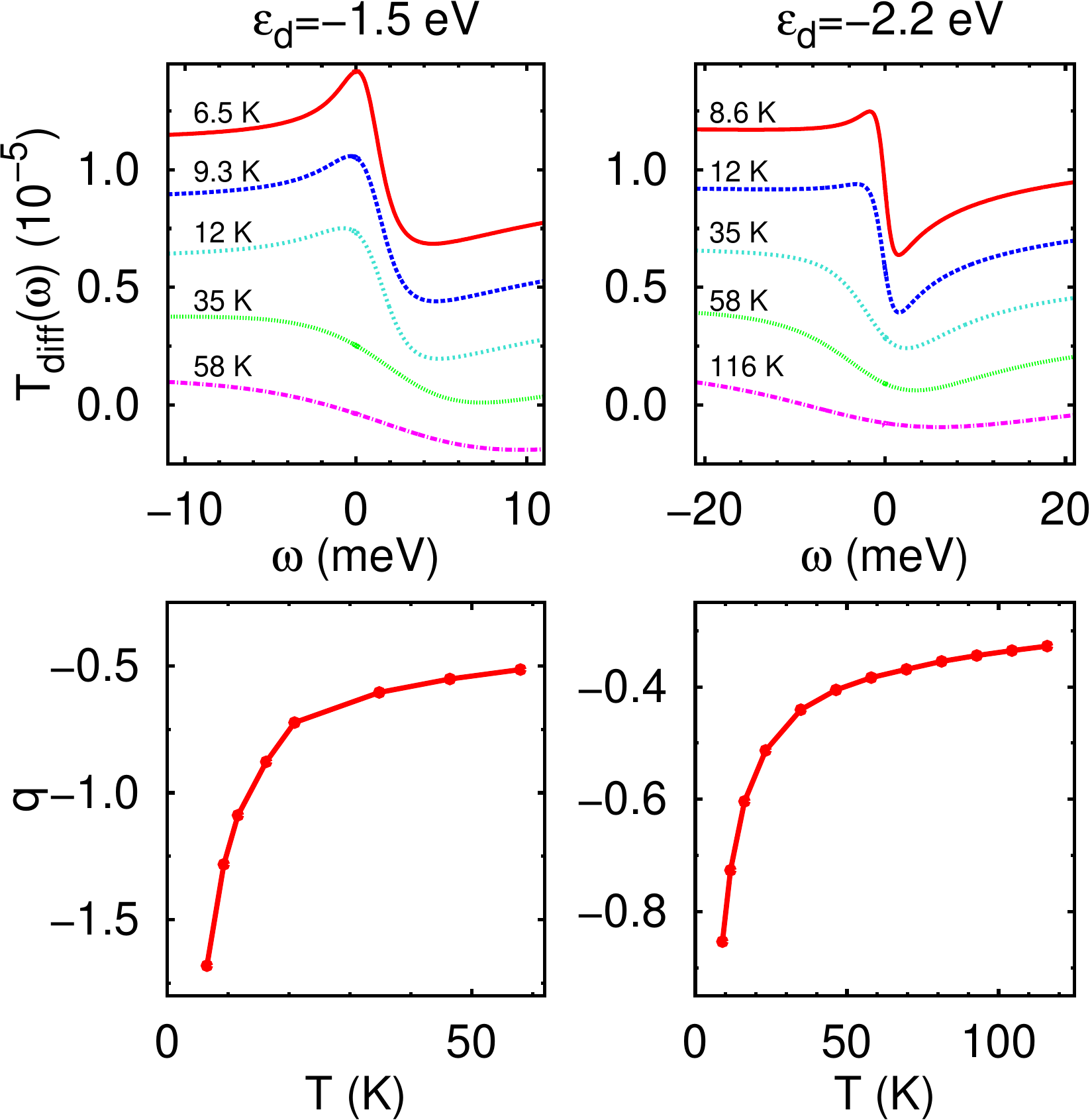}
   \caption{Temperature dependence of two different line shapes for Co$@$Cu(111), 
     $xz$, $U=3$~eV, $\epsilon_d=-1.5$~eV and $\epsilon_d=-2.2$~eV, 
     respectively. Top: Transmission. Bottom: $q$ parameter; the lines are a guide for the eye. }
   \label{fig_T_dependence}
  \end{center}
 \end{figure}

The results presented so far are for the case of $T\rightarrow 0$ (according to the criterion discussed in 
Sec.~\ref{sec_results_001}). We now study the temperature dependence of two line shapes: One tending towards 
a peak ($q>1$) and one tending towards a dip ($q<1$). We pick the $xz$ orbital of Co$@${}Cu(111), $U=3$~eV, 
$\epsilon_d=-1.5$~eV ($q=-1.49$) and $\epsilon_d=-2.2$~eV ($q=-0.87$), respectively. 
The top row of Fig.~\ref{fig_T_dependence} shows the evolution of the aforementioned two line shapes. 
For increasing temperature, the signal amplitude diminishes, while its width grows. The peak does not decay 
symmetrically. The 'peak' component of the Fano feature decays faster than the 'dip' component of the feature, 
so that, in both cases, the feature as a whole becomes increasingly dip-like with increasing temperature. 
In order to quantify that, we perform Fano-Frota fits and calculate the $q$ parameter. We find that the 
$q$ parameter decreases considerably when temperature is rising, 
irrespective if the feature tends more towards peak or dip in the 
$T\rightarrow 0$ case. 
\section{A simplified model}

\label{sec_models}
\begin{figure}
  \begin{center}                         
   \includegraphics[width=.6\linewidth]{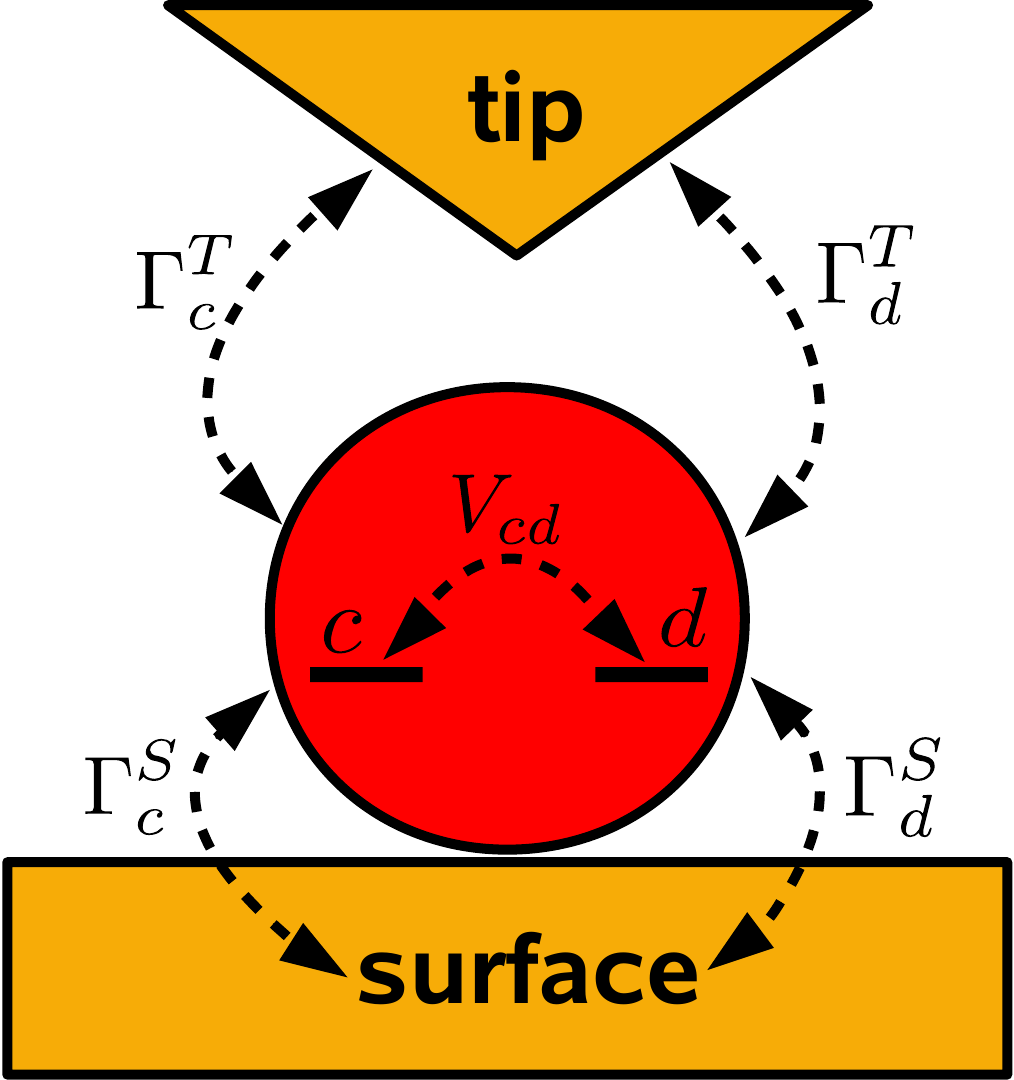}
   \caption{Sketch of the simplified model. The effective atom A is described by the correlated $d$-level and one conduction 
     electron level $c$, in contact with the surface and the tip. 
   \label{fig_model}}
  \end{center}
 \end{figure}

The interference mechanism leading to different Fano line shapes still is a matter of discussion 
\cite{Ujsaghy2000,Schiller2000,Plihal2001,Madhavan2001,Wahl2004,Luo2004,Merino2004a,Merino2004b,Luo2004,Lin2006,Zitko2010}. 
We expand on this discussion by introducing a simple model that allows us to 
determine transmission line shapes from \emph{ab initio} parameters. Fig.~\ref{fig_model} shows a schematic drawing of 
our model system. The central assumption is that the quantum interference primarily occurs \textit{on} 
the magnetic adatom, namely between one $s$-type and/or $p$-type level (in the following, we will simply 
call it the conduction level $c$) and the correlated $d$-level. Both levels are in contact to the tip T and the surface S, and 
the respective interactions are taken into account by coupling matrices $\Gamma_{\rm{T/S}}$. As a second 
central assumption we neglect the direct tunneling from the tip to the surface. 

The starting point of our model is the correlated Green's function of the effective atom
comprising the conduction $c$-level and the correlated $d$-level of the magnetic atom. 
\begin{eqnarray}
  \label{eqn_M1}
  \lefteqn{G_{\rm{A}}(\omega) = \left( \omega \hat{P}_{\rm A}- \hat{H}_{\rm A} -\hat{\Delta}(\omega) - \Sigma_d(\omega)\hat{P}_d \right)^{-1}} \nonumber\\ 
  &&= \begin{pmatrix} \omega -\epsilon_c - \Delta_c(\omega) & \quad -V_{cd} - \Delta_{cd}(\omega) \\
    -V_{dc} - \Delta_{dc}(\omega) & \quad \omega - \epsilon_d - \Delta_{d}(\omega) - \Sigma_d(\omega) 
  \end{pmatrix}^{-1} 
  \nonumber\\                      
\end{eqnarray}

$\hat{P}_{\rm A}$ is a projector onto the effective atom A, while $\hat{P}_d$ projects onto the $d$-level only.
All parameters can either be extracted from the KS-calculation ($\epsilon_d$, $\epsilon_c$, $V_{cd}$, $\hat{\Delta}(\omega)$) or 
from the OCA-calculation ($\Sigma_d(\omega)$), while the chemical potential has been set to zero $\mu = 0$.
The diagonal elements of the hybridization function $\hat{\Delta}(\omega)$
lead to a shift (real part) of the level position of $\epsilon_c$ and $\epsilon_d$, respectively, and yield an effective level 
broadening (imaginary part). Also note that the hybridization function has off-diagonal components 
$\Delta_{cd}(\omega) = \Delta_{dc}(\omega)$, which can be understood as an additional hopping between 
the $c$- and $d$-level mediated by hoppings via the substrate,
to give a total effective coupling of $\tilde{V}_{cd} = V_{cd} + \Delta_{cd}$. 
The coupling matrices $\Gamma_{\rm{S/T}}(\omega)$ necessary for calculating the transmission function
by Eq.~\ref{eqn_caroli} can be obtained by decomposing the hybridization function into 
a tip ($\Delta_{\rm{T}}(\omega)$) and a surface ($\Delta_{\rm{S}}(\omega)$) component and taking the 
imaginary parts, i.e. $\Gamma_{\rm{S/T}}(\omega) = -2\,\Im\,\Delta_{\rm{S/T}}(\omega)$. 

For the conduction level $c$ of the effective atom we choose the $s$- or $p$-orbital that couples to the correlated 
$d$-orbital. In the case of the (001) and the (111) substrates the $z^2$-orbital couples to the $s$- as 
well as the $p_z$-orbital. In this case we apply a unitary transformation in the subspace of the $s$- and 
$p_z$-orbitals such that the $z^2$-orbital decouples completely from one of the orbitals in the new basis. 
The $sp_z$-hybridized orbital coupling to the $z^2$ is then found to be the linear combination 
$\ket{sp_z} \propto \tilde{V}_{sz^2}\ket{s}+\tilde{V}_{p_zz^2}\ket{p_z}$ where $\tilde{V}_{sz^2}$ and 
$\tilde{V}_{p_zz^2}$ are the effective hoppings of the $z^2$-orbital with the $s$- and $p_z$-orbitals, respectively.
On both surfaces, the $xz$-orbital couples to $p_x$ and the $yz$-orbital to $p_y$.
For the (001) surface both the $x^2-y^2$- and the $xy$-orbitals 
do not interact with any of the $s$- or $p$-orbitals on the atom, while on the (111) surface, they do interact 
with the $p_y$- and $p_x$-orbitals, respectively.

\begin{figure}
  \begin{center}                         
   \includegraphics[width=\linewidth]{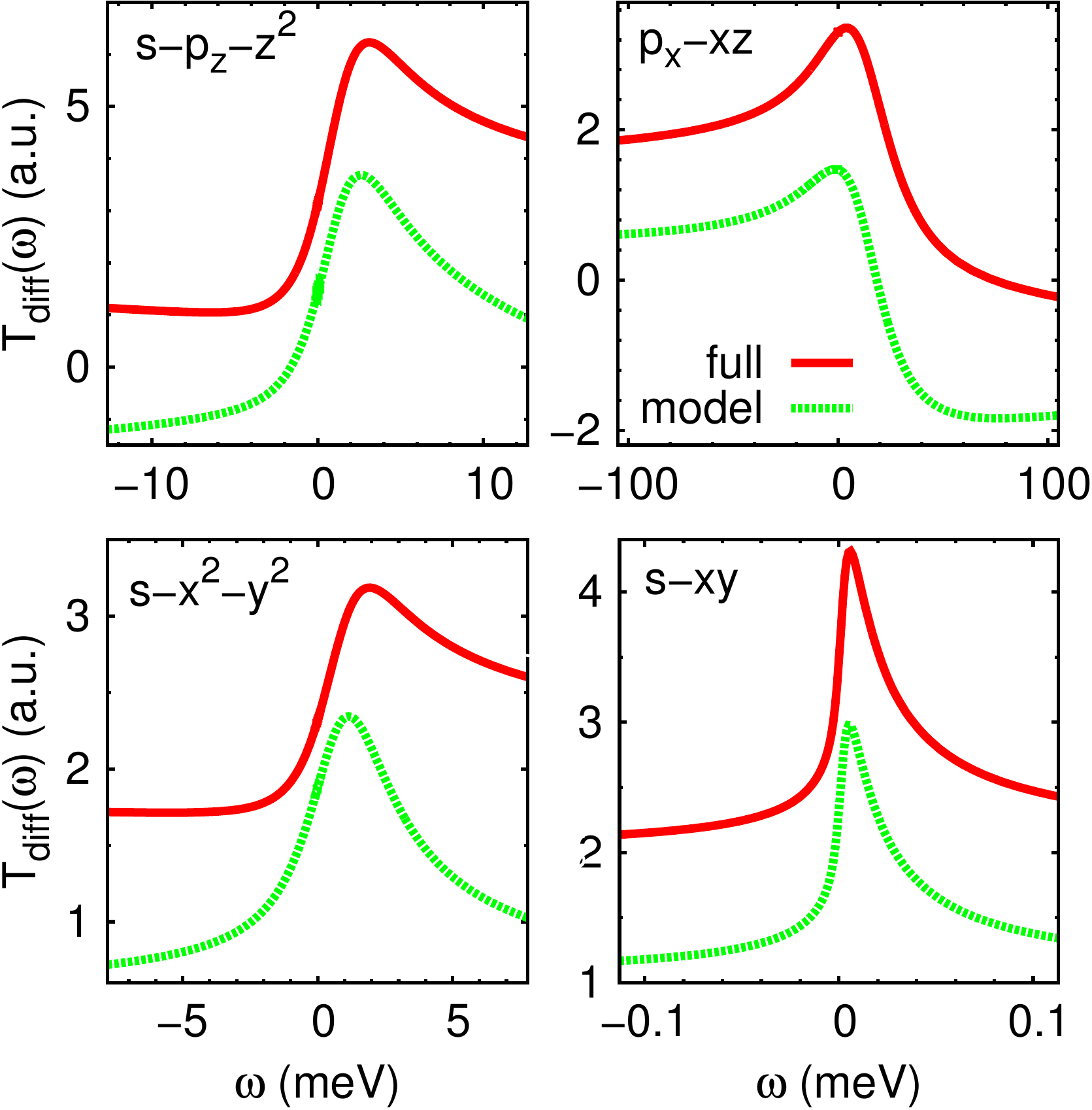}
   \caption{Transmissions calculated \emph{ab initio} with the ANT.G package (see Sec.~\ref{sec_results}) and for the simplified 
   model; Co$@$Cu(001), $U=2$~eV, $\epsilon_d=-1$~eV. The transmission functions are rescaled and offset for better visibility.}
   \label{fig_model_001}
  \end{center}
\end{figure}

\begin{figure}
  \begin{center}                         
    \includegraphics[width=\linewidth]{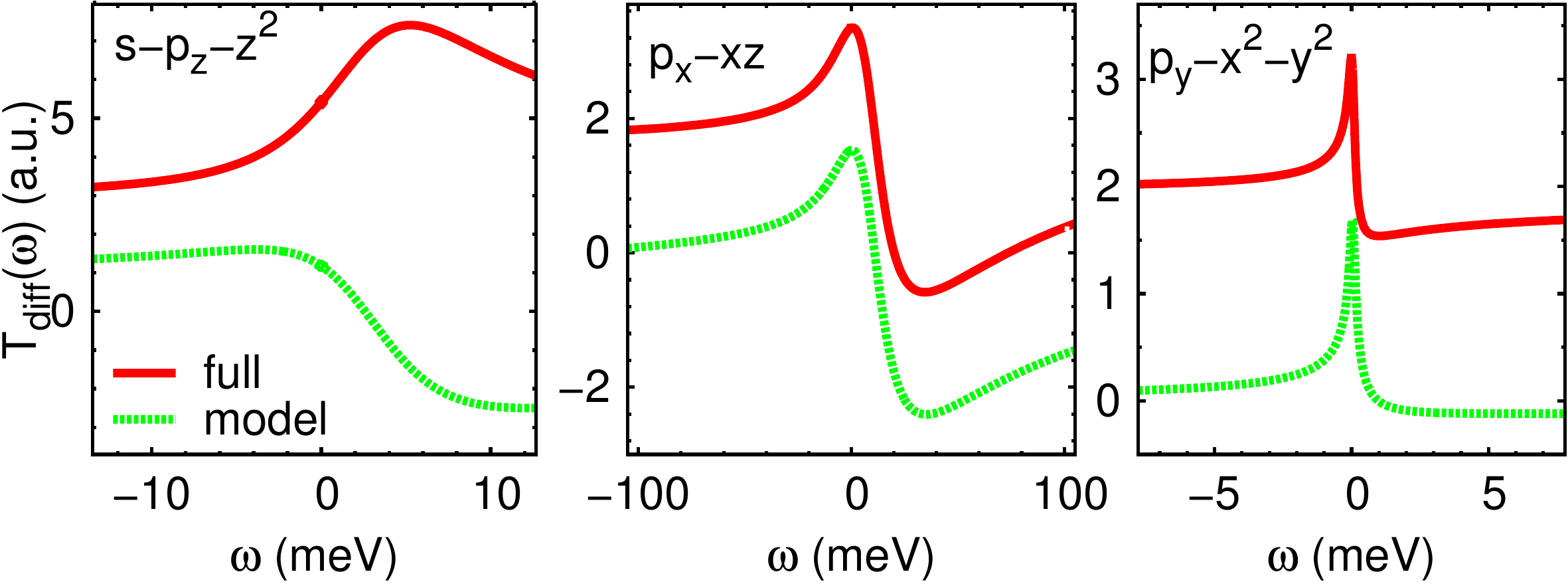}
    \caption{Transmissions calculated \emph{ab initio} with the ANT.G package (see Sec.~\ref{sec_results}) and for the simplified 
    model; Co$@$Cu(111), $U=2$~eV, $\epsilon_d=-1$~eV. The transmission functions are rescaled and offset for better visibility.
    }
    \label{fig_model_111}
  \end{center}
\end{figure}

In Fig.~\ref{fig_model_001} and \ref{fig_model_111}, we compare line shapes calculated for the simplified model 
with the full \emph{ab initio} results from Sec.~\ref{sec_results}. 
For the Co$@$Cu(001) surface, the simplified model consisting of the $z^2$-orbital and the $sp_z$-hybridized 
orbital reproduces the line shape of the $z^2$-orbital quite well. Only the peak character is slightly overestimated. 
In the case of the $xz$-orbital the line shape of the simplified model including the $p_x$-orbital is in excellent
agreement with that of the full \emph{ab initio} calculation. For the $x^2-y^2$-orbital the agreement between the simplified
model and the full calculation is not as good. As stated before this orbital does not interact with any $s$- or $p$-orbital
on the Co atom. Hence the transmission of the simplified model reproduces simply the Kondo peak in the spectral
function since no interference is taking place. On the other hand the full transmission shows a somewhat asymmetric Fano 
feature ($q\approx1.7$) indicating that interference with some substrate state(s) must take place, which is not included in the model. 
Finally, for the $xy$-orbital we find very good agreement between the simplified model and the full calculation.
The line shape in both calculations simply reproduces the Kondo peak in the spectral function of the $xy$-orbital
indicating the absence of any interference effects between this $d$-level and $s$- and $p$-levels on the atoms
as well as substrate states.

We find a somewhat similar picture for Co$@$Cu(111). For the $xz$-orbital the model including the interaction
with the $p_x$-orbital gives a line shape in excellent agreement with the full calculation. Also for the $x^2-y^2$ 
the simplified model including the $p_y$-orbital on the atom reproduces the line shape of the full calculation
very well. However, in the case of the $z^2$-orbital the simplified model including the $sp_z$-hybridized
orbital fails quite badly in reproducing the line shape of the full calculation. Apparently, interference with
tunneling paths to substrate states play an important role here.

\section{Discussion}
\label{sec_discussion}

For Co$@$Cu(001), we found transmission line shapes ranging from asymmetric Fano features with positive 
($z^2$, $x^2-y^2$) and negative ($xz$) $q$ values to a more peak-like feature ($xy$). The line shapes are 
determined by the interference of different tunneling paths. Our simplified model calculations indicate 
that for $z^2$ and $xz$ the interference takes place on the adatom between the correlated $d$-level
and the non-interacting $sp$-levels coupling to the $d$-orbital.
For the $xy$-orbital, no interference occurs between the conduction and impurity tunneling 
channels. Hence one directly observes the shape of the Kondo peak in the transmission. 
On the other hand, for the $x^2-y^2$-orbital, the interference mechanism probably involves the 
Cu substrate states which are not captured by the simplified model. 
Experimentally, asymmetric Fano line shapes were reported with $q\sim1.1-1.2$ in the tunneling 
regime\cite{Knorr2002,Neel2007}. The measured line shapes are comparable to the features we found 
both in the $z^2$- and $x^2-y^2$-orbitals (see Figs.~\ref{fig_001_U2} and \ref{fig_001_U3}), although 
the $z^2$-orbital yields a slightly better agreement. Better agreement with experiment can surely
be achieved by adjusting the Anderson model parameters and fitting the calculated spectra with
the experimental ones. We would like to stress though that finding good agreement with experiment 
is not the primary goal of this work, but rather to demonstrate how different orbital symmetries
give rise to different Fano-Kondo line shapes.
A recent study by one of us\cite{Jacob2015} found an underscreened Kondo effect for Co$@$Cu(001), 
where the $z^2$ and $x^2-y^2$ are nearly half filled, but only the $z^2$ orbital is Kondo screened at 
finite temperatures due to its higher Kondo temperature. Ref.~\onlinecite{Baruselli2015} 
comes to similar conclusions, finding a Kondo peak in the $z^2$-orbital with $q=1.2$ 
in the tunneling regime and explaining it due to the interference of the $z^2$- with the $s$-orbital.

For Co$@$Cu(111), we found asymmetric to peak-like Fano line shapes with positive ($z^2$) and negative 
($xz$, $x^2-y^2$) $q$ values. For the latter two, we can understand the tunneling interference in terms 
of the model presented in the previous section. The interference occurs on the magnetic atom, between 
the conduction electron channel, modeled by one of the $p$-orbitals, and the respective $d$-level. 
For $z^2$, which is interacting with the $sp_z$ hybridized level, our model fails, indicating that 
interference with substrate states plays an important role here.

Experimentally, dips were reported with $q$ values close to zero\cite{Knorr2002,Manoharan2000, Vitali2008}
which does not seem to agree with any of the calculated line shapes. The $z^2$-orbital, aligned in the transport 
direction, again shows the strongest signal, but is rather peak-like. The closest candidate to a dip-like line shape 
is the $xz$-orbital, particularly when increasing the occupancy relative to half-filling by moving the $d$-level position 
downwards in energy (see Fig.~\ref{fig_111_U3}). 
In Sec.~\ref{sec_results_Tdep}, we studied its temperature dependence, and found that the line shape 
became increasingly dip-like when increasing temperature. However, note that in our calculations for the $xz$-orbitals 
we find $q<0$ while in experiment $q$ is always positive.

Probably, the surface state of the Cu(111) surface\cite{Crommie1993} plays an 
important role for determining the line shape~\cite{Manoharan2000,Bogicevic2000,Merino2004b,Lin2006}
since its tunneling amplitude may be twice as strong compared to tunneling into bulk states\cite{Jeandupeux1999}.
However, our embedded cluster calculation probably does not capture the surface state properly.
The importance of the surface state for reproducing the correct line shape in the Co$@$Cu(111) 
system is also stressed in Ref.~\onlinecite{Baruselli2015} where the surface state is not properly
captured and the correct $q$ value could not be reproduced either.

\section{Conclusions}
\label{sec_conclusions}

In summary, we have calculated the orbital signatures of Kondo peaks in the STM spectra of 
transition metal adatom systems, namely Co$@$Cu(001) and Co$@$Cu(111). Our calculations show
that the measured line shapes allow us to draw some conclusions on the $d$-orbital(s) involved 
in the Kondo effect since the line shape depends to a large extent on the coupling of the 
$d$-orbital to the $sp$-orbitals on the adatom, which in turn is determined by the orbital 
symmetry. However, also temperature, effective interaction $U$ and in particular the occupancy
of the $d$-orbital have a strong influence on the actual line shapes. Also, if multi-orbital 
effects are important for the actual shape of a Fano-Kondo feature, this approach per se is 
not appropriate. Nevertheless, even in the case of a multi-orbital Kondo effect, often one 
orbital will be dominant in the tunnel spectra. In fact if a Kondo resonance forms in the 
$z^2$-orbital, the corresponding Fano feature will be dominant in the tunnel spectrum for 
the typical case of an $s$-type STM tip, so that Kondo features coming from other $d$-orbitals 
are likely not visible. These results are also relevant for STS of transition metal complexes 
on metallic substrates \cite{Jacob2013, Kuegel2014, Karan2015}, maybe even more so since tunneling 
into surface states is less important there. 

We stress that the here developed method %%of investigating the orbital nature of Fano-Kondo line shapes 
can in principle also be applied to the contact regime. However, unlike in the tunneling case, in
the contact regime the voltage can no longer be assumed to mainly drop between tip and adatom. 
Rather, the voltage drop will distribute in some way over the contact according to the actual
geometry of the contact region~\cite{Karan2015}, and needs to be calculated or estimated. 
Moreover, the actual contact geometry is probably also relevant for the coupling between 
$d$-orbitals and conduction electrons and thus also has a strong influence on the line shapes. 
Therefore possible contact geometries need to be explored and relaxed with some care.

Based on our results, we propose a poor man's method to obtain information on the orbital(s)
involved in the Kondo effect measured in an actual experiment solely on the basis of a density 
functional theory calculation of the system: by tailoring an appropriate self-energy for each
orbital such that the width of the resulting Kondo peak in that orbital reproduces the width
of the measured Fano-Kondo line shape, one can calculate the corresponding line shapes and compare
to experiment. 

\section*{Acknowledgments}

We acknowledge fruitful discussions with R. Requist, M. Karolak  and J. J. Palacios.

\begin{appendix}

\section{Complex and real Fano line shapes}
\label{appendix_A}

Here we derive the real Fano line shape given by Eq.~(\ref{eqn_fano_alternative}) 
from its complex representation in Eq.~(\ref{eqn_fano_lorentz}):
\begin{equation}
  \rho_{\rm FL}(\omega) =  \Im \left[ e^{i \phi_q} \left( \frac{A}{\omega - \omega_0 + i \Gamma} \right) \right] 
\end{equation}
Introducing the abbreviation $\epsilon = (\omega-\omega_0)/\Gamma$, we have
\begin{eqnarray*}  
  \rho_{\rm FL} &=& \frac{A}{\Gamma} \cdot \Im \left[ \left(\cos( \phi_q) + i \sin(\phi_q)\right)  \frac{\epsilon-i}{\epsilon^2 + 1 }  \right] \\
  &=& \frac{A}{\Gamma} \left[ \frac{-\cos(\phi_q)+\epsilon \sin(\phi_q)}{\epsilon^2+1} \right] \\
  &=& \frac{A}{\Gamma} \left[\frac{-\cos^2(\frac{\phi_q}{2}) + \sin^2(\frac{\phi_q}{2}) +2\epsilon \sin(\frac{\phi_q}{2}) 
      \cos(\frac{\phi_q}{2})}{\epsilon^2+1} \right] \\
  &=& \frac{A}{\Gamma} \left[\frac{-1+\tan^2(\frac{\phi_q}{2})+2\epsilon \tan({\frac{\phi_q}{2}})}{\epsilon^2+1} \right]
  \cos^2\left(\frac{\phi_q}{2}\right) 
\end{eqnarray*}

Defining $q \equiv \tan(\phi_q/2)$, we arrive at
\begin{equation}
 \rho_{\rm FL}  = \frac{A}{\Gamma} \left[ \frac{(q+\epsilon)^2}{\epsilon^2+1} -1 \right] \frac{1}{1+q^2}
\end{equation}
which is the same as Eq. (\ref{eqn_fano_alternative}).

\end{appendix}

\bibliography{PhDBib}

\end{document}